\documentclass[namedreferences]{SolarPhysics}
\usepackage[optionalrh]{spr-sola-addons} 
\usepackage{graphicx}        
\usepackage{color}           
\usepackage{url}             
\usepackage[breaklinks=true]{hyperref}

\newcommand{\etal}{{\it et al.}}

\newcommand{\apj}{    {\it Astrophys. J.}}

\newcommand{\solphys}{{\it Solar Phys.}}

\begin{document}

\begin{article}

\begin{opening}

\title{Features of spatial distribution of oscillations in faculae regions\\ {\it Solar Physics Topical Issue on helio- and astero-seismology}}

\author{N.I.~\surname{Kobanov}$^{1}$ and
        V.A.~\surname{Pulyaev}$^{1}$\sep       }
\institute{Institute of Solar-Terrestrial Physics, Irkutsk, 664033, P.O. Box
291,  Russia (e-mail: kobanov@iszf.irk.ru)}
\runningauthor{Kobanov et al.}
\runningtitle{Facula oscillations}

\begin{abstract}
We found that oscillations of LOS velocity in H$\alpha$ are different
for various parts of faculae regions. Power spectra show that the
contribution of low-frequency modes (1.2\,--\,2 mHz) increase at
the network boundaries. Three- and five- minute periods
dominate inside cells. The spectra of photosphere and chromosphere LOS velocity
oscillations differ for most faculae. On the other hand, we detected several
cases where propagating  oscillations in faculae were manifest
with a five-minute period. Their initiation point on
spatial-temporal diagrams coincided with the local maximum
of the longitudinal magnetic field.

\end{abstract}
\end{opening}

\section{Introduction}
     \label{S-Introduction}
The heating of the solar corona is an old problem in solar physics.
It is intended that waves propagating from the photosphere into the
chromosphere and further into the corona are responsible for energy transport. Solar facula structures are quite widespread on
the Sun, they are present even in polar areas that are not accessible
to sunspots. Usually they are observed as associations of bright merged
areas. They regularly occupy large regions of the solar surface and can
play a significant role in processes of upward energy transport. In
some papers, photosphere oscillations in the faculae are considered
as a source of five-minute oscillations observed in the upper
chromosphere, transition zone and corona \cite{DePont03,Cent06}. The behaviour of facula oscillations has
been actively studied since 1960s \cite{Orr65,How67,Shee71,Deub74}. Later, investigations were continued by \inlinecite{Wo81}, \inlinecite{Balt90}, \inlinecite{Mug95}, \inlinecite{Khom08}, \inlinecite{Cent09}. Nevertheless, the problem is still far from solution.
The aim of this paper, as a continuation of our previous work (Kobanov and Pulyaev
2007, hereafter Paper I), is to is to explore the behaviour of chromosphere oscillations in  various parts of faculae regions.

\section{Instrument and method}
  \label{S-Instrument and method}
Observational material was obtained over several years on the horizontal
solar telescope of the Sayan Solar Observatory, located at an altitude of 2 km
on a mountain peak. The telescope is supported six meters from the Earth
surface and equipped with a wind-shield system. The diameter of the coelostat
mirrors is 800 mm, and the focal length of the main mirror is 21 m. A photoelectric
guiding system provides image tracking with an accuracy  of about 1 arcsec,
with compensation for the excursion of the image due to the Sun rotation.
The observer can change orientation of the observed object relative to the
spectrograph entrance slit using a Dove prism. A Princeton Instrument
CCD camera (256 x 1024) was used. One pixel along the entrance slit corresponds to
0.24$^{\prime\prime}$ and along the spectrograph dispersion to 6\,--\,8 m{\AA}.
Because of seeing the average spatial resolution was about 1.5\,--\,2$^{\prime\prime}$.
We made observations in the H$\alpha$\,6562.8\,{\AA}, Fe\,{\small I}\,6569.2\,{\AA},
Ca\,{\small II}\,8542\,{\AA} and Fe\,{\small I}\,8538\,{\AA} spectral lines.
Sometimes we used polarization optics to measure the strength of the longitudinal
magnetic field simultaneously with the line-of-sight velocity. In order to reduce
the influence of the aliasing effect the measurement process was organized such
that the time interval between exposures was shorter than the exposure time. We
applied corrections for tilt in the spectra, the influence of dirt on the CCD,
and the flat field. Velocity was calculated using the center-of-gravity method at a range of $\pm$0.03{\AA} from the spectral line center for photospheric lines, $\pm$0.15{\AA} for H$\alpha$, $\pm$0.1{\AA} for Ca {\small II}. Shift of the center-of-gravity point correspond to line-of-sight (LOS) velocity.  The longitudinal magnetic field was calculated from the observed Zeeman splitting, determined as  the distance between the centers of gravity of the polarized components. We then obtained gray-scale spatial-temporal diagrams of
the LOS velocity, intensity, and magnetic-field strength (for observations with the polarization optics).

\section{Results and Discussion}
 \label{S-Results and Discussion}
\begin{figure}[]
 \centerline{\includegraphics[width=0.9\textwidth]{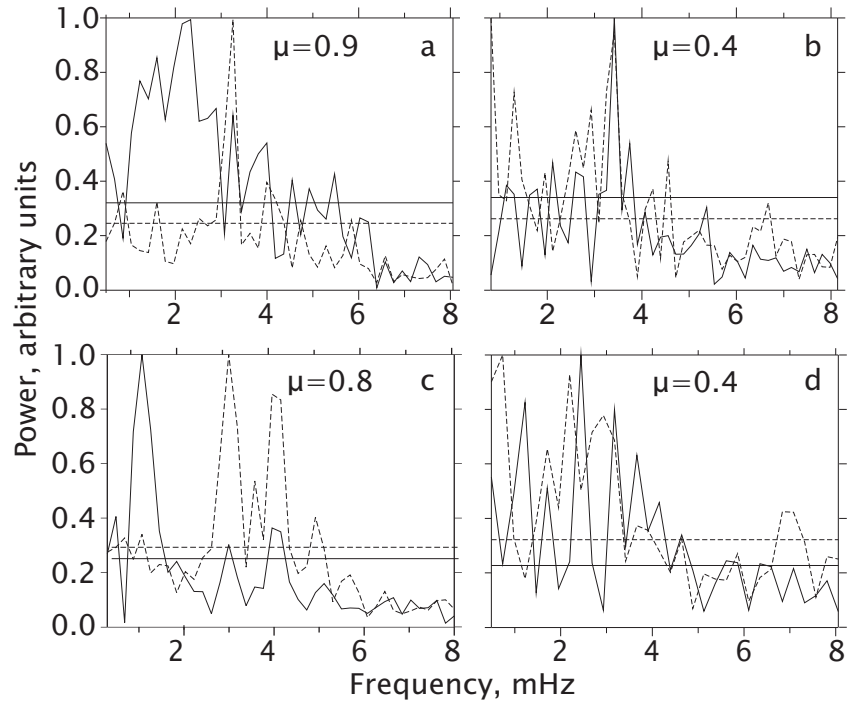}}
    \caption[Figure 1]{Examples of LOS velocity power spectra  for some faculae, averaged over 2 arcs. Solid line - chromosphere, dashed - photosphere. a) in spectral lines: Ca\,{\small II}\,8542\,{\AA}, Fe\,{\small I}\,8538\,{\AA}. b,c,d) in spectral lines: H$\alpha$\,6563\,{\AA}, Fe\,{\small I}\,6569\,{\AA}. Solid and dashed horizontal lines indicate of 95\% confidence levels, $\mu=\cos\theta$}
 \label{fig:1}
\end{figure}

In our observations, the locations of facular areas near the limb were determined using white-light images (D=175mm) of the guide system. Sometimes, (for faculae far from limb) we used slit-jaw images in H$\alpha$ or Ca\,{\small II}\,K. To obtain more accurate links with fine-structure facular elements, we used bright intervals of the spectrum. The coincidence of power spectra of photospheric and chromospheric oscillations may be considered as evidence of the fact, that five-minute photosphere oscillations penetrate into the chromosphere. We have analyzed 32 time series for 32 faculae. The average duration of time series is about an hour. Cadence varied from 1 to 10 seconds. However, comparison of oscillation spectra carried out in the H$\alpha$\,6562.8 {\AA},  Fe\,{\small I} 6569.2 {\AA} , Ca\,{\small II} 8542 {\AA} and Fe\,{\small I} 8538 {\AA} spectral lines more often reveals differences than likeness (Figure~\ref{fig:1}). One can assume that the projection effect is a reason of revealed discrepancy. It is known, the contribution of low frequency oscillations can increase towards the limb \cite{Stix74}. However, two power spectra  in figure 1 ($\mu$=0.8 - 0.9) show the same differences.

\begin{figure}[]
 \centerline{\includegraphics[width=0.9\textwidth]{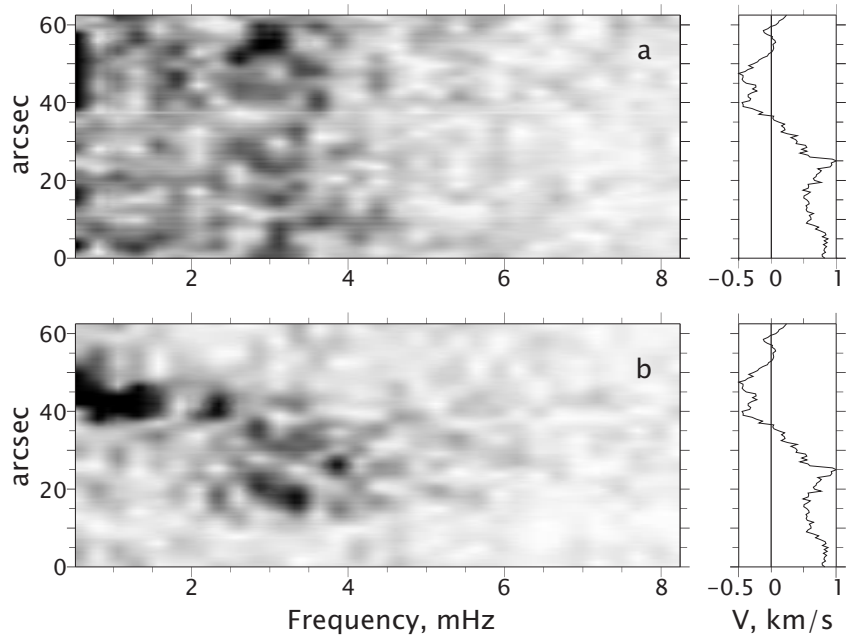}}
    \caption[Figure 2]{The space localization of different frequency modes for facula area ($\mu$=0.77), top-photosphere, bottom-chromosphere. Darkest elements indicate strongest oscillations. On the right - space slice of chromosphere LOS velocity averaged on the whole time series, the same space slice see in Figure 3, bottom.}
 \label{fig:2}
\end{figure}

\begin{figure}[]
 \centerline{\includegraphics[width=0.9\textwidth]{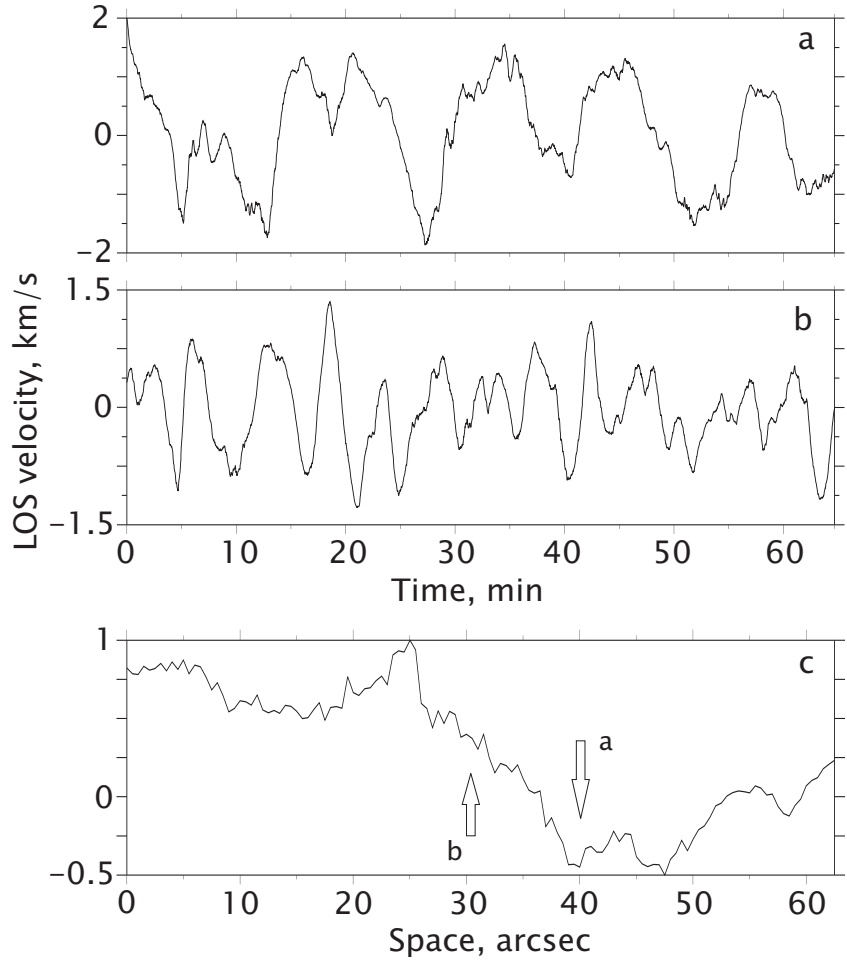}}
    \caption[Figure 3]{Unfiltered LOS velocity signals in different parts of the facula (same as Fig.\,2). a)The low-frequency mode at point 40 arcsec (arrow a in bottom panel). b) five minute oscillation at point 30 arcsec (arrow b). c) Space slice of LOS velocity averaged on the whole time series.}
 \label{fig:3}
\end{figure}
\begin{figure}[]
  \centerline{\includegraphics[width=0.9\textwidth]{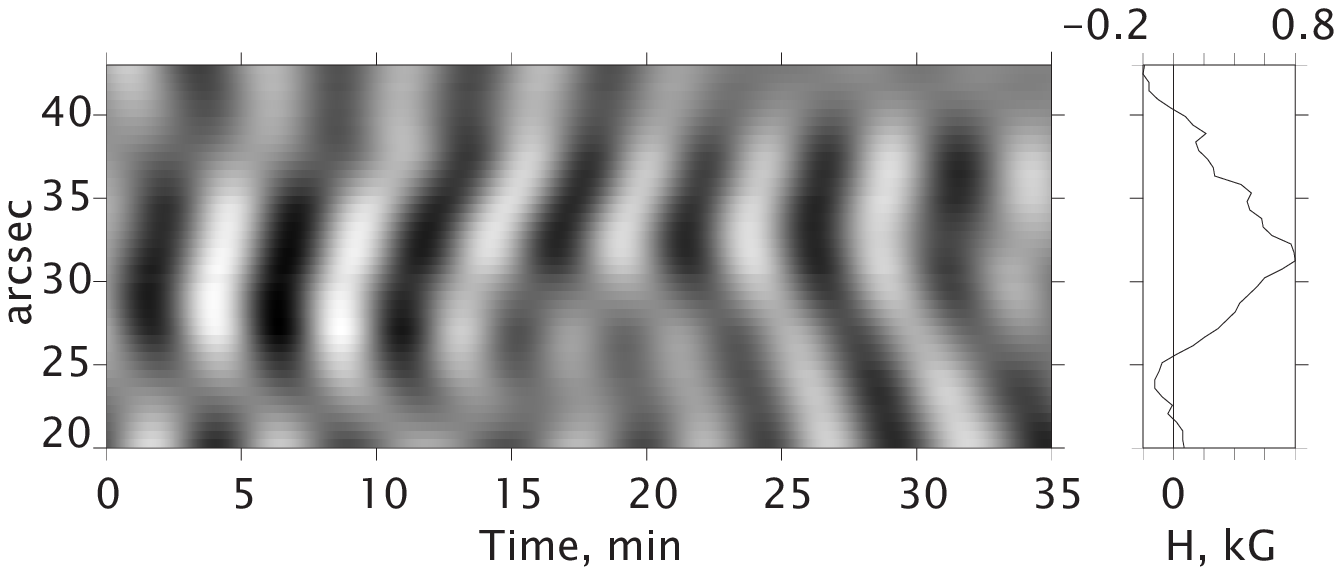}}
      \caption[Figure 4]{The traveling five-minute waves in chromosphere of the facula (the same facula as Figs 2,3). The dark parts indicate LOS velocity away from observer. On the right - space slice of longitudinal magnetic field.}
  \label{fig:4}
\end{figure}
 We conclude that the differences are not a consequence of instrumental effects. Our result is contrary to one obtained by \citeauthor{Cent06} \shortcite{Cent06, Cent09}, \inlinecite{Khom08}. They made observations (in  He\,{\small I}\,10830 {\AA}, Si\,{\small I}\,10827\,{\AA}) of a facula area located near the disk center on June 14, 2004.  The illustration given in these papers shows spectra of photosphere and chromosphere oscillations averaged on 40 arcsec. The coincidence of the spectra is impressive.
Currently we have no idea about the origin of the above mentioned contradictions. Possibly the reason lies in the use of different spectral lines and  different ranges of heights observed in the solar atmosphere accordingly. Early \inlinecite{Liv76} established that the formation height of He\,{\small I}\,10830 {\AA} line is significantly reduced over a facula area. Taking in account that the level for the Si\,{\small I}\,10827 {\AA} line is estimated as 540 km \cite{Bard08}, one can assume convergence of the observed levels above faculae. The latter is no more than a speculation to be tested by more observations in He\,{\small I} and Si\,{\small I}.

 Even  if the same frequencies are present in the space-averaged photosphere and chromosphere spectra considerable differences are registered in gray-scale diagrams showing space localization of the power of these oscillations (Figure~\ref{fig:2}). It can be seen that the different frequency modes belong to spatial elements of a facula remote from each other.

 The reason for the differences could be the local topology of magnetic fields in different parts of a facula. It is known, that the chromosphere network is not destroyed in facula areas. We identify the net boundaries as facula parts where a quasi-steady down flow is observed. For that, LOS velocity signal is averaged on the whole time series.
Figure~\ref{fig:3} shows unfiltered signals of a chromospheric LOS velocity in  individual elements of the same facula region as in Figure~\ref{fig:2}. Low-frequency oscillations (Figure~\ref{fig:3}a) dominate in the area we define as a chromosphere network boundary in accordance with the above rule. Five-minute oscillations (Fig.~\ref{fig:3}b) clearly appear in the region where the quasi-stationary LOS velocity is close to zero. This area is 10 arcsec away from the network boundary.
 In spite of the above problems, the propagating waves in the chromosphere of faculae are observed on the 3 mHz frequency (Paper I).
 We applied a frequency-filtration method \cite{Kob04} for the construction of gray-scale space-time diagrams of the Doppler velocity for various frequency modes, to  better distinguish propagating waves and establish their parameters. These diagrams reflect the dynamics of the Doppler velocity  along a spatial direction, determined by the position of the image on the spectrograph entrance slit. The resulting gray-scale images of space-time distributions of the LOS velocity in the H$\alpha$\ line show a clear periodic structure resembling a chevron. The chevron directly demonstrates the presence of traveling waves in the facula chromosphere (Figure 4). The peak of chevron on the space-time diagram coincided with the local maximum of the longitudinal magnetic field in the photosphere. It is likely that this point is a loop base, where the magnetic field is almost vertical. The horizontal phase velocity determined from the chromospheric chevron slope is equal to 50\,--\,70  km/s. We have failed to find a clear phase relationship  between chromosphere waves and LOS velocity oscillations in the photosphere of faculae.

\section{Conclusion}
 \label{S-Conclusion}

The coincidence of the spectral composition of  LOS velocity oscillations in the
photosphere and chromosphere of faculae, considered as direct evidence of links between these
layers, is observed relatively rarely. Moreover, even in cases of similarity of mean power spectra,
the space-time localization of the power of the modes analyzed do not coincide.
 We found that the frequency content of LOS velocity oscillations is non-homogeneous in different parts of facula areas.  The power of low-frequency modes (1.2\,--\,2 mHz) increase at network boundaries. Three- and five- minute periods dominate inside cells. We observed some cases where  propagating waves in the chromosphere above facula were manifest with a five-minute period. Their initiation point on spatial-temporal diagrams coincided with the local maximum of the longitudinal magnetic field.


\begin{acks}
 This work was partially supported by  the Program of State Support for Leading Scientific Schools of the Russian Federation (NSh-3552.2010.2), the Federal Agency for Science and Innovation (State contract 02.740.11.0576) and Basic Research Program No.\,16 (part 3) of the Presidium of the Russian Academy of Sciences. We are very grateful to the anonymous referee, whose valuable remarks, comments and suggestions helped to improve this paper.
\end{acks}


\IfFileExists{\jobname.bbl}{} {\typeout{}
\typeout{****************************************************}
\typeout{****************************************************}
\typeout{** Please run "bibtex \jobname" to obtain} \typeout{**
the bibliography and then re-run LaTeX} \typeout{** twice to fix
the references !}
\typeout{****************************************************}
\typeout{****************************************************}
\typeout{}}

\end{article}

\end{document}